\documentclass[preprint,double spaced]{revtex4}%
\usepackage{amsfonts}
\usepackage{amsmath}
\usepackage{amssymb}
\usepackage{graphicx}%
\begin{document}
\title{Giant Nonlinear Electron-lattice Interaction in Cuprate Superconductors, and
Origin of the Pseudogap}
\author{R. Nistor$^{2}$, G.J. Martyna$^{1}$, M. H. M\"{u}ser$^{1,2}$, D.M. Newns$^{1}%
$, and C.C. Tsuei $^{1}$}
\affiliation{$^{1}$IBM T.J. Watson Research Center, Yorktown Heights, NY 10598, USA}
\affiliation{$^{2}$Dept. of Applied Mathematics, University of Western Ontario, London,
Ontario, CANADA N6A 5B7}

\begin{abstract}
The pseudogap is a key property of the cuprate superconductors, whose
understanding should illuminate the pairing mechanism. Recent experimental
data support a close connection between the pseudogap and an oxygen-driven C4
symmetry breaking within the CuO$_{2}$ plane unit cell. Using ab initio
Molecular Dynamics, we demonstrate the existence of a strong nonlinear
electron-oxygen vibrator coupling in two cuprates. In a mean field approach
applied to this coupling within a model Hamiltonian, we derive a C4
splitting/pseudogap phase diagram in agreement with experiment - providing an
explanation for the pseudogap phenomenon from first principles. The
implications for superconductivity and the Fermi surface arc effect are discussed.

\end{abstract}
\startpage{1}
\endpage{ }
\maketitle

In 20 years no consensus has emerged on the mechanism for $d$-wave high
temperature superconductivity (HTS) in cuprates or for their characteristic
pseudogap (PG) \cite{PGReview}. \ Alternatives such as the electron-lattice
coupling which mediates pairing in conventional superconductors
\cite{Maksimov}, and purely electronic mechanisms \cite{PWA,Yang}, have been
widely discussed. A possibly decisive input comes from recent work
\cite{Kohsaka1,Kohsaka2} showing a strong correlation between the occurrence
of the PG in the underdoped region and the breaking of C4 symmetry (i.e. the
$a$ and $b$ directions become nonequivalent) in the planar oxygen vibrational
amplitudes. This observation points to the origin of the PG as a lattice
instability, and is a very important clue - consistent with evidence such as
the doping-dependent oxygen isotope shift \cite{alpha1,alpha2} and the
HTS-induced softening in oxygen vibration frequency
\cite{Ramanshift,neutronshift} - that electron-lattice coupling is the
mechanism for HTS. Paradoxically, \textit{ab initio} calculations do not find
a strong conventional \textit{linear} electron-lattice coupling in the
cuprates \cite{Marvin,Bohnen}, leading to the proposition that a primarily
\textit{nonlinear} coupling lies at the heart of HTS.

In this paper we use \textit{ab initio} molecular dynamics (AIMD) to show that
electron coupling to the vibrations of O in the CuO$_{2}$ plane is in fact
very strong - but it is \textit{second order} in the oxygen displacement, in
contradiction with the assumptions underlying the earlier electron-lattice
coupling work \cite{Marvin,Bohnen}. To explain why the coupling should be
second order, we argue from symmetry that the local effect on the electrons of
a radial displacement of oxygen in the Cu-O-Cu bond should be independent of
the displacement's sign, and hence second order in the O displacement. Based
on this symmetry argument, two of us earlier put forward the Fluctuating Bond
Model of HTS \cite{FBM1} which is empirically based on the idea of second
order coupling, an idea which had been proposed earlier
\cite{Bussman,Crespi,Mahan}. The FBM successfully predicted C4 symmetry
breaking \cite{FBM1} and enabled superconducting properties such as $d$-wave
symmetry \cite{Changreview} to be understood. Unlike conventional
superconductivity, the FBM has not yet been parametrized from first
principles, and here we show that the value of the FBM coupling energy does
indeed have the value required in the FBM \cite{FBM1}, validating its key
assumption of a giant nonlinear electron-lattice coupling.

The technique of \textit{ab initio} molecular dynamics (AIMD) \cite{AIMD}
solves the ionic equations of motion on a Born-Oppenheimer potential energy
surface obtained by solution of the many-electron Schrodinger equation in
local density approximation, a technique well suited to the present problem in
that it avoids the\textit{ }constraint of a linearized interaction.%
\begin{figure}
[h]
\begin{center}
\includegraphics[
trim=0.654452in 1.452500in 1.757976in 1.319124in,
height=2.0652in,
width=3.1557in
]%
{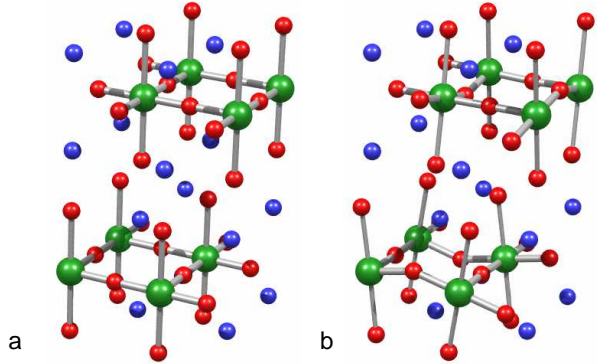}%
\caption{\textit{ab initio} Molecular Dynamics calculation at
$T=4\operatorname{K}$ of the structure of metallic La$_{2}$CuO$_{4}$ ($214$)
\cite{Bianconi}, blue spheres, La, green spheres, Cu, red spheres, O.
\textbf{a}, Undistorted setup structure, \textbf{b}, Equilibrated structure
showing vertical displacements of the planar oxygens correspond to rotations
of CuO$_{6}$ octahedra about alternate $x$- and $y$- axes in planes stacked
along the $c$-axis - the LTT structure found at low temperature in metallic
$214$ phases.}%
\end{center}
\end{figure}
We start by applying AIMD to the C4 symmetry-breaking in two cuprates. First
we consider the well-established Low Temperature Tetragonal (LTT) structure
\cite{Bianconi} found in underdoped metallic 214 materials, which was early on
linked to non-linear electron lattice coupling \cite{Pickett}. Figure 1 shows
the LTT structure is predicted correctly by AIMD for the metallic La$_{2}%
$CuO$_{4}$ at $T=4%
\operatorname{K}%
$. All CuO$_{2}$ planes undergo C4 symmetry-breaking in which one half of the
oxygens undergo a $z$-displacement, the $\pi/2$ rotation of successive planes
around the $c$-axis ensures overall tetragonal symmetry. The symmetry breaking
by $z$-displacement in the 214 family contrasts with the symmetry breaking by
$xy$-displacement in the 2212 and oxychloride (Ca$_{2}$CuO$_{2}$Cl$_{2}$)
materials \cite{Kohsaka1}.%
\begin{figure}
[hh]
\begin{center}
\includegraphics[
trim=1.001365in 0.501788in 1.067343in 0.000000in,
height=3.0796in,
width=3.4575in
]%
{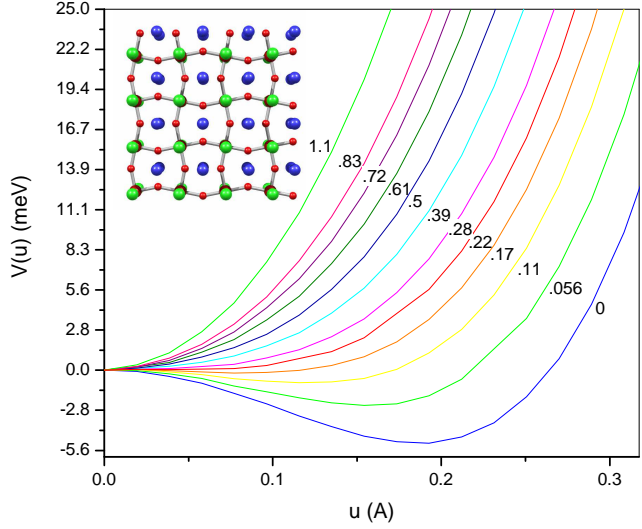}%
\caption{PE curves $V(u)$ for oxychloride material as a function of oxygen
in-plane distortion $u$ (see insert) for different dopings (see labelling on
curves). Doping is implemented by fractional substitution of Na for Ca.}%
\end{center}
\end{figure}

We now establish a model-independent global connection between oxygen PE
surface and the number of electrons by extending the Fig. 1 type of
calculation to a range of dopings. The oxychloride system Ca$_{2-x}$Na$_{x}%
$CuO$_{2}$Cl$_{2}$ is computationally attractive and doping can be controlled
via the Na fraction $x$, the calculated oxygen PE surfaces being shown in Fig.
2. AIMD accurately reflects experiment \cite{Kohsaka1} in that the oxygen
displacemements are in the $xy$-plane (Fig. 2 inset) (but cannot capture its
non-Born-Oppenheimer features). The PE surfaces in Fig. 2 are strongly
dependent on doping, their minimum is off the bond axis (leading to the Fig. 2
inset distortion) at low doping, transitioning to an on-axis\ minimum
(stability of oxygen on the bond axis) at high doping, and can be parametrized
in the form%
\begin{equation}
V(u)=\left(  \chi+Vp\right)  \frac{u^{2}}{2}+\frac{w}{8}u^{4}.
\label{PE_surface}%
\end{equation}
Here $u$ is oxygen displacement from the bond axis, the zero-doping bond force
constant $\chi$\ is negative (oxygen unstable at zero doping), and the
electron-lattice coupling $V$ is positive, representing stabilization of the
unstable Cu-O-Cu bond with increasing doping $p$. The positive quartic term
$w$\ represents the confinement of oxygen in the local lattice cage. The
results of Fig. 2 and Eq. (\ref{PE_surface}) clearly show that the oxygen
potential energy surface is strongly dependent on global electron number, with
a coupling term in $V$ going as the\textit{ square }of the oxygen
displacement, as was argued above. If the Cu-O-Cu bond is sensitive to the
global number of electrons, we can also anticipate a strong response of the
oxygen to the \textit{local} electron number.

To calculate the local form of electron-lattice coupling requires a model. We
start from the Emery tight-binding model based on Cu $3d_{x^{2}-y^{2}}$ and
the oxygen $2p_{x}$/$2p_{y}$ orbitals that have $\sigma$ symmetry in the
Cu-O-Cu bond. The coupling is determined by considering how the matrix
elements vary with oxygen displacement, and projecting the result back into a
one-band basis of just the $3d_{x^{2}-y^{2}}$ orbitals (see Supp. Mat.). The
resulting complete FBM\ Hamiltonian Eq. (\ref{FBM}) is written in mixed
representation electronically. It considers only one vibrational mode per
oxygen for simplicity:%
\begin{align}
H  &  =\sum_{\mathbf{k},\sigma}\epsilon_{\mathbf{k}}n_{\mathbf{k},\sigma}%
+\sum_{\left\langle i,j\right\rangle }\left[  \frac{p_{ij}^{2}}{2M}+\frac
{\chi_{0}}{2}u_{ij}^{2}\right]  +\frac{w}{8}\sum_{\left\langle
i,j\right\rangle }u_{ij}^{4}\label{FBM}\\
&  -\frac{v}{\sqrt{2}}\sum_{\left\langle i,j\right\rangle }N_{ij}u_{ij}%
^{2}.\nonumber
\end{align}
Equation (\ref{FBM}) embodies the conventional 1-band model \cite{OKAPRL} with
a single $3d_{x^{2}-y^{2}}$ orbital per Cu atom located at site $i$ in the
CuO$_{2}$ plane. Here $c_{i,\sigma}^{+}$ ($c_{i,\sigma}$) is the creation
(destruction) operator for the $3d_{x^{2}-y^{2}}$ orbital on site $i$ of spin
$\sigma$, with number operators $n_{i,\sigma}=c_{i,\sigma}^{+}c_{i,\sigma}$,
and $c_{\mathbf{k},\sigma}^{+}$ ($c_{\mathbf{k},\sigma}$) are the
corresponding operators for the band states of wavevector $\mathbf{k}$, etc..
A sum over $\left\langle ij\right\rangle $ implies that each nearest-neighbor
bond $ij$ appears only once in the sum. The variables $p_{ij}$, $u_{ij}$ are
the conjugate momentum and position coordinates of oxygen in bond $ij$, $M$ is
oxygen mass and $\chi_{0}$ the bare oxygen force constant. $w$ is the quartic
interaction, and $v$ a (positive) electron-vibrator coupling constant. The
operator $N_{ij}=\sum_{\sigma}\left(  n_{i\sigma}+n_{j\sigma}-c_{i,\sigma}%
^{+}c_{j,\sigma}-c_{j,\sigma}^{+}c_{i,\sigma}\right)  $ represents the number
of electrons in the $ij$-antibonding orbital $\left[  c_{i,\sigma}%
-c_{j,\sigma}\right]  /\sqrt{2}$.

In more detail the first term of Eq.(\ref{FBM}) is the electronic band energy,
and the second term represents the harmonic part of the oxygen Hamiltonian, to
which is added the third term, a quartic interaction which confines the oxygen
and is needed to fit the PE curves in Fig 2. The nonlinear electron-vibrator
coupling is represented in the fourth term, which involves the vibrator
coordinate $u_{ij}$ quadratically, and the electronic operator $N_{ij}$.
Dumping electrons $N_{ij}$ into the antibonding orbital in bond $ij$ softens
the vibrator $ij$, as implied by Eq. (\ref{PE_surface}) and by the trend in
Fig. 2 - it is also reasonable on intuitive chemical grounds. Note that the
$c_{i,\sigma}^{+}c_{j,\sigma}+c_{j,\sigma}^{+}c_{i,\sigma}$ term in $N_{ij}$,
which originates from processes where an electron hops from Cu $i$ to O and
then to Cu $j$, is identical to the original FBM \cite{FBM1}, while the
present analysis adds a new term $-\left(  n_{i\sigma}+n_{j\sigma}\right)  $,
originating from processes where an electron hops from Cu to O and then ends
up back on the same Cu.

The FBM coupling is strong between the oxygens and the saddle points at
X$=\left(  \pi,0\right)  $ and Y$=\left(  0,\pi\right)  $ in the band
structure - the energies at X and Y are normally degenerate, but the
degeneracy is split by C4 symmetry-breaking. The bare electron-lattice
coupling constant $v$ in Eq. (\ref{FBM}) can be simply determined by
displacing the $x$-oxygens and determining the shift in the band structure
eigenvalue at X. Any effect of a chemical potential shift can be removed by
displacing the $y$-oxygens and subtracting the two effects (see Supp. Mat.).
The results are collected in Table I for the 214 and oxychloride materials.
The values of the quartic interaction $w$ are obtained from the quartic
coefficient of the fit to the Fig. 2 curves, and similar ones for the 214
material. It is found that the coupling $v$ is relatively small for
vibrational polarization along the Cu-O-Cu bond, so we only considered
polarizations transverse to the bond in Table I.\bigskip

\begin{center}
{\small Table I: FBM Parameters for 214 and Oxychloride Materials}%

\begin{tabular}
[c]{|l||l|l||l|l|}\hline
Polarization & $v_{214}$ $\left(  \mathrm{au}\right)  $ & $v_{\mathrm{oxy}}$
$\left(  \mathrm{au}\right)  $ & $w_{214}$ $\left(  \mathrm{au}\right)  $ &
$w_{\mathrm{oxy}}$ $\left(  \mathrm{au}\right)  $\\\hline\hline
$xy$ $\bot$ to bond & $0.0163$ & $0.0182$ & $0.053$ & $0.090$\\\hline
$z$ $\bot$ to bond & $0.0174$ & $0.0202$ & $0.122$ & $0.106$\\\hline
\end{tabular}

\end{center}

Note that the coupling strengths $v$ in Table I are rather material- and
polarization- independent, while the anharmonic coefficents $w$ vary by less
than $2.5\times$.

The pairing energy in the FBM is second order in the coupling $v$, with the
functional form \cite{FBM1} $v^{2}/w$. In the current form Eq.(\ref{FBM}) it
is $K^{\prime}=4v^{2}/w$ (see Supp. Mat.). From Table I $K^{\prime}$ can be
estimated to be about $K^{\prime}\simeq0.5$ $%
\operatorname{eV}%
$, matching the giant magnitude required \cite{FBM1} for high temperature
superconductivity. This is the first meaningful \textit{ab initio} prediction
of the electron-lattice coupling constant in HTS, and points the way to an
eventual quantitative theory of HTS\ by extending the existing body of work
describing conventional superconductors.

Here, as an application of the model Eq.(\ref{FBM}), rather than again
presenting results for the gap equation \cite{FBM1} which will be done
elsewhere, we present an important new result, the pseudogap phase diagram.
This is done by the mean field decoupling of Eq.(\ref{FBM}) (see Supp. Mat.),
(a) $N_{ij}u_{ij}^{2}\rightarrow\left\langle N_{ij}\right\rangle u_{ij}^{2}$ ,
leading to a softening of the vibrator proportional to the number of
antibonding electrons $\left\langle N_{ij}\right\rangle $, and (b)
$N_{ij}u_{ij}^{2}\rightarrow N_{ij}\left\langle u_{ij}^{2}\right\rangle $
wherein the nearest-neighbor hopping $\left(  c_{i,\sigma}^{+}c_{j,\sigma
}+c_{j,\sigma}^{+}c_{i,\sigma}\right)  $ is reduced by increased vibrator
amplitude $\left\langle u_{ij}^{2}\right\rangle $. In this mean field
approximation the model decouples into exactly soluble band structure and
anharmonic oscillator problems, linked by the mean field quantities
$\left\langle N_{ij}\right\rangle $ and $\left\langle u_{ij}^{2}\right\rangle
$, which are solved for self-consistently assuming they are uniform in space.

When the expectation values $\left\langle N_{ij}\right\rangle $ and
$\left\langle u_{ij}^{2}\right\rangle $ in the $x$- and $y$-directed bonds are
found to differ, C4 symmetry breaking in both \textit{vibrator amplitudes} and
the \textit{electronic structure }occurs\textit{.} Symmetry breaking in the
oxygen vibrator amplitudes is exactly the effect measured in the STM
experiments of Ref. \cite{Kohsaka1} (consistently we predict that the
higher/lower-amplitude oxygens in the Ref. \cite{Kohsaka1} R-plots should be
dark/light streaks (see Supp. Mat.), as observed). C4 symmetry breaking in the
electronic structure, wherein the degeneracy of the saddle points at
X$=\left(  \pi,0\right)  $ and Y$=\left(  0,\pi\right)  $ is split, leads to a
$d$-type PG $\Delta_{ps}\left(  \mathbf{k}\right)  \sim\Delta_{ps}\left(  \cos
k_{x}-\cos k_{y}\right)  /2$ (see Supp. Mat.), where the PG, $\Delta_{ps}$,
can be positive or negative in sign. The PG $\Delta_{ps}$ is one half the
splitting between the energies at X and Y. The work of Ref. \cite{Kohsaka2}
shows that the experimental PG is intimately associated with the C4 splitting
- confirming the correctness of identifying $\Delta_{ps}$ with the
experimental pseudogap. The predicted PG phase diagram is shown in Fig. 3, and
is seen to reproduce the known experimental phase diagram for the pseudogap
\cite{ParisRaman,ARPES1,ARPES2,Kohsaka2,PGReview}. At low temperature the
pseudogap ranges from a maximum of about $100$meV on the underdoped side,
decreasing with increasing doping
\cite{ParisRaman,ARPES1,ARPES2,Kohsaka2,JinhoLee}, while $T^{\ast}$ behaves
reasonably as a function of doping and is of the correct magnitude
\cite{PGReview}.%

\begin{figure}
[h]
\begin{center}
\includegraphics[
trim=1.383395in 1.108487in 2.819998in 0.841735in,
height=3.1185in,
width=3.2474in
]%
{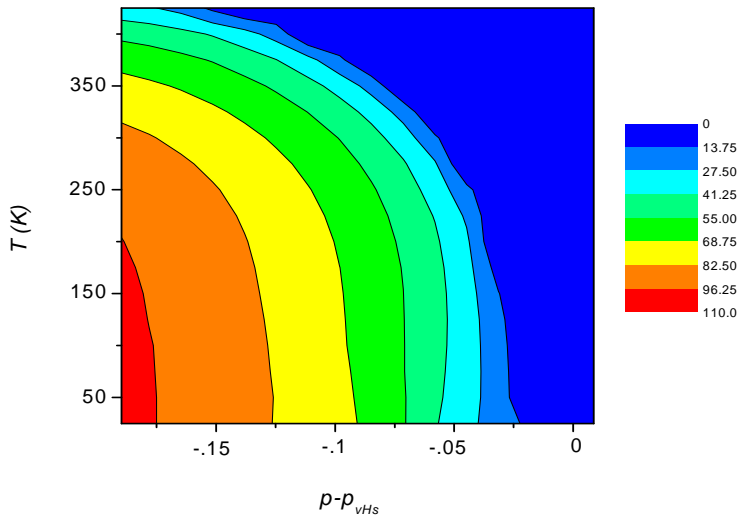}%
\caption{Contour map of pseudogap in temperature/doping plane showing decrease
with doping, and with temperature, until it vanishes at phase boundary
$T^{\ast}$. Contours labelled by pseudogap $\Delta_{ps}$ in intervals of
$13.75$ $\mathrm{meV}$. For experimental $\Delta_{ps}$ magnitudes see Ref.
\cite{JinhoLee}.}%
\end{center}
\end{figure}
Note that experimentally, the C4 splitting is characterized by a nanoscale
domain structure \cite{Kohsaka1}. In our picture this implies that the sign
of\ $\Delta_{ps}$ will vary spatially, in a manner such as $\Delta_{ps}\left(
\mathbf{k}\right)  \sim\cos\theta\Delta_{ps}\left(  \cos k_{x}-\cos
k_{y}\right)  /2$, where $\theta$ is some local phase determined by the local
nanoscopic disorder. The Fermi surface in the C4 split phase will be sensitive
to the sign of $\Delta_{ps}$. The inset in Fig. 4 shows the two FS
corresponding to the extreme cases $\theta=0$ and $\theta=\pi$, and it is
reasonable to assume that a nanoscopically varying order parameter will lead
to FS smearing as illustrated by the shaded region in the Fig. 4 inset.

The smearing of the FS by the PG, seen in the Fig. 4 inset, is seen to be
least near the nodal line $k_{x}=k_{y}$, a manifestation of the $d$-type
nature of the PG $\Delta_{ps}\left(  \mathbf{k}\right)  $, and to increase as
one goes towards the SP's at X and Y. If we make a measurement on some energy
scale $E$, it might be expected that the FS will be well-defined at
$\mathbf{k}$-points where the local PG is less than $E$, $\Delta_{ps}\left(
\mathbf{k}\right)  <E$, but that it will appear smeared out on energy scales
where the local PG is larger than $E$, $\Delta_{ps}\left(  \mathbf{k}\right)
>E$. As an example of this phenomenon we could take $E$ to be temperature,
$E=k_{\mathrm{B}}T$ . According to this argument, there should then be a
boundary between the resolvable and unresolvable sections of the FS arc
defined by $\Delta_{ps}\left(  \mathbf{k}\right)  =k_{\mathrm{B}}T$. In fact
this FS arc effect \cite{Norman} has already been observed a couple of years
ago, as shown by a comparison of the data with our heuristic model in Fig. 4,
which is in good agreement with experiment \cite{Kanigel} (this approach does
not include the temperature-dependent quasiparticle lifetime broadening
\cite{Norman}, which we hope to derive in a future paper dealing with
quasiparticle dynamics).%

\begin{figure}
[t]
\begin{center}
\includegraphics[
trim=0.600181in 0.360279in 1.255697in 0.369225in,
height=2.9888in,
width=3.5405in
]%
{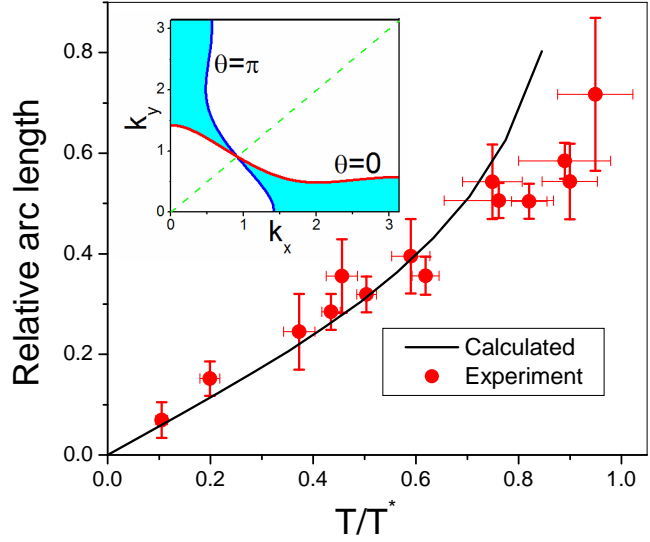}%
\caption{Inset: two FS's for domains with opposite sign of pseudogsap
($\Delta_{ps}=74$ $\mathrm{meV}$), colored area indicates approximate loss of
definition of FS in a nanoscopically disordered sample. Plot of FS arc length
vs. temperature compared with experiment (see text). }%
\end{center}
\end{figure}

The approach here has been to focus on the electron-lattice interaction while
ignoring the strong electron correlation effects. It would be productive to
consider the nonlinear electron-lattice coupling in the context of a strongly
correlated (large-$U$) system, where the modulation of the Cu-Cu bond by
oxygen vibration is expected to be as strong as in the noninteracting case,
and the phase diagram can be extended to include magnetic effects.

In conclusion, using AIMD we have demonstrated the instability of the
CuO$_{2}$ plane oxygens in three cuprates which relates closely to experiments
such as the LTT structure in the 214 family and the C4 symmetry breaking in
the 2212 and oxychloride materials. The nonlinear electron-lattice coupling
deduced from these calculations is large with the magnitude required by the
FBM theory. Using our deduced Hamiltonian Eq. (2), we are able to derive the
pseudogap phase diagram, harmonizing the pseudogap data with the observed C4
splitting, and to understand the temperature-dependent Fermi arc observations.

\end{document}


\title{Giant Nonlinear Electron-lattice Interaction in Cuprate Superconductors, and
Origin of the Pseudogap}
\author{R. Nistor$^{2}$, G.J. Martyna$^{1}$, M. H. M\"{u}ser$^{1,2}$, D.M. Newns$^{1}%
$and C.C. Tsuei $^{1}$}
\affiliation{$^{1}$IBM T.J. Watson Research Center, Yorktown Heights, NY 10598, USA}
\affiliation{$^{2}$Dept. of Applied Mathematics, University of Western Ontario, London,
Ontario, Canada N6A 5B7}
\startpage{1}
\maketitle

\section{\textit{ab initio} Molecular Dynamics}

The Car-Parrinello ab initio MD \cite{AIMD} studies presented herein were
performed using Kohn-Sham Density Functional Theory. The CPMD approach is
based on the Born-Oppenheimer approximation. It describes the metallic state,
but not the magnetic, insulating state, of the cuprate materials. The
computations were checked by comparison with the band structures of copper and
the 214 material \cite{Pickett2}. The Fig. 1 results were obtained by starting
with the Fig. 1a configuration and running to equilibrium. The Fig. 2b results
were obtained by constraining the oxygens in the manner shown in Fig. 1a at a
specified value of the oxygen displacement, and running to equilibrium - they
therefore represent "ion-relaxed" potential energy curves. Doping was achieved
by varying the K/Ca ratio in the ensemble. Samples of 4x4 and 6x6 unit cells
were run in periodic boundary conditions.

\section{Projection of 3-band model onto 1-band model}

In matrix notation consider a $d$-subspace and a $p$-subspace, represented by
the Hamiltonians $H^{d}$ and $H^{p}$ respectively, connected by the coupling
matrix $V^{pd}$, the Hamiltonian then being%

\begin{equation}
H=\left[
\begin{array}
[c]{cc}%
H^{d} & V^{dp}\\
V^{pd} & H^{p}%
\end{array}
\right]  .
\end{equation}
Projecting onto the $d$-subspace in perturbation theory%

\begin{equation}
\widetilde{H^{d}}=H^{d}+V^{dp}\left(  \epsilon_{d}-H^{p}\right)  ^{-1}V^{pd}.
\end{equation}
if $i,j$ are $d$-sites, and $l,m$ are p-orbitals%
\begin{equation}
\widetilde{H^{d}}_{ij}=\epsilon_{d}\delta_{ij}+\sum_{l,m}V_{il}^{dp}\left(
\epsilon_{d}-H^{p}\right)  _{lm}^{-1}V_{mj}^{pd}.
\end{equation}
%

\begin{figure}
[h]
\begin{center}
\includegraphics[
trim=0.526754in 0.526999in 0.529947in 0.792938in,
height=2.6161in,
width=3.6703in
]%
{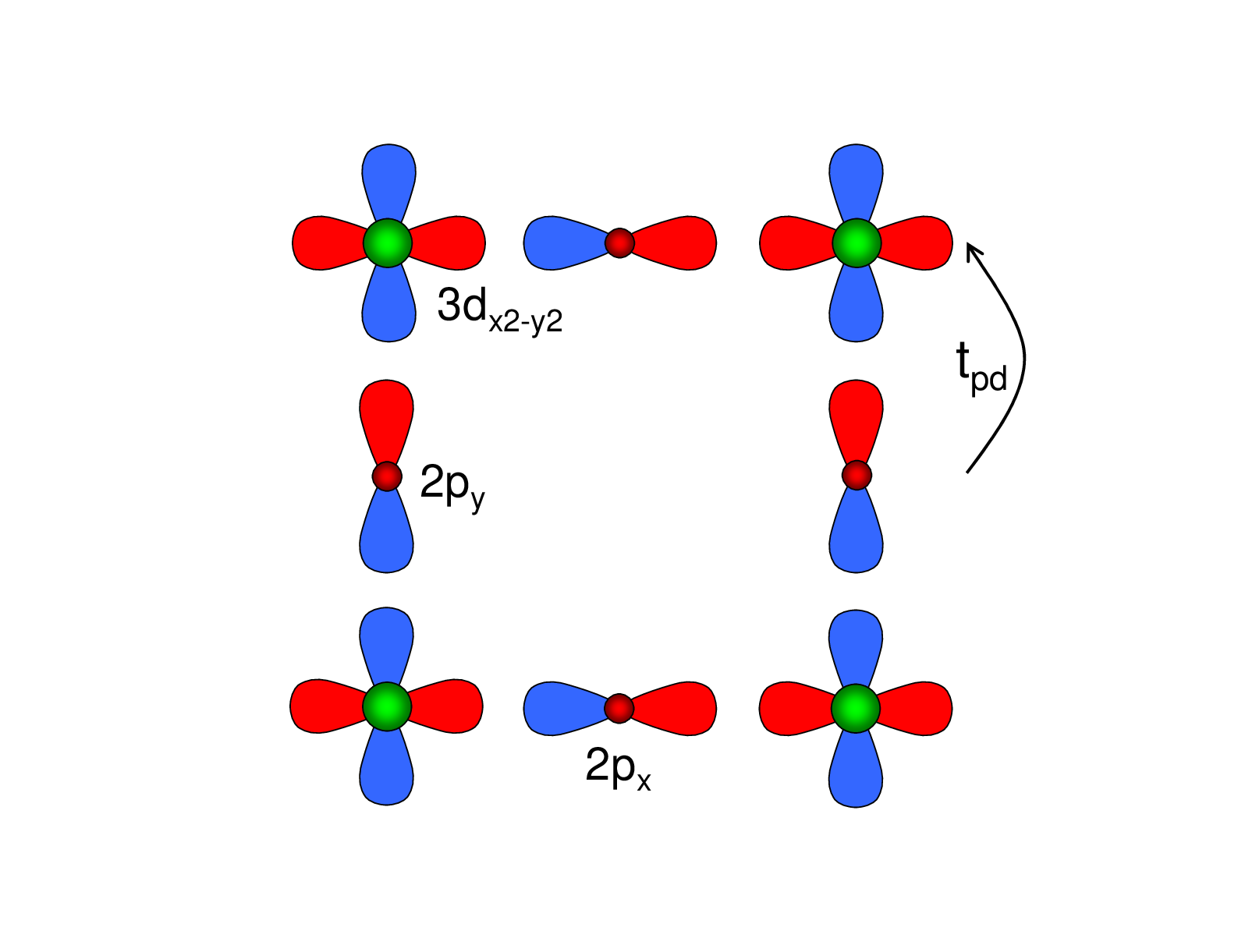}%
\caption{$2p_{x}$, $2p_{y}$ and $3d_{x^{2}-y^{2}}$ orbitals in CuO$_{2}$
plane, illustrating $2p$ to $3d$ hopping integral $t_{pd}$.}%
\end{center}
\end{figure}

Now we shall neglect the $pp$ hopping matrix elements (Emery model), when
$l=m$ and the $V$'s are nearest-neighbor hopping matrix elements defined as
$t_{pd}>0$ (see Fig. 1). There are 2 processes,

\begin{enumerate}
\item $i,j$ nearest neighbor $\left\langle i,j\right\rangle $ on the
$d$-lattice, when the 2 $V$'s have opposite sign (Fig. 1)

\item $i=j$, when the 2 $V$'s have same sign
\end{enumerate}

giving%
\begin{equation}
\widetilde{H^{d}}=\sum_{i,\sigma}\epsilon_{d}n_{i\sigma}+\sum_{\left\langle
i,j\right\rangle ,\sigma}\frac{t_{p_{ij}d}^{2}}{\epsilon_{p_{ij}d}}\left(
n_{i\sigma}+n_{j\sigma}\right)  -\sum_{\left\langle i,j\right\rangle }%
\frac{t_{p_{ij}d}^{2}}{\epsilon_{p_{ij}d}}X_{ij}, \label{formulation1}%
\end{equation}
where $\epsilon_{pd}=\epsilon_{d}-\epsilon_{p}>0$ is the "oxide gap" between
the oxygen $2p$ orbital energy and the higher-lying Cu $3d_{x^{2}-y^{2}}$
orbital energy, $\sigma$ is spin, $p_{ij}$ is the p-orbital between $d$-sites
$i$ and $j$, and the bond order operator $X_{ij}$\ is
\begin{equation}
X_{ij}=\sum_{\sigma}\left(  c_{i\sigma}^{+}c_{j\sigma}+c_{j\sigma}%
^{+}c_{i\sigma}\right)  .
\end{equation}

Let us assume that the oxygen motion in some direction is $x$, and that it
enters the $3$-band hamiltonian via the $pd$ hopping integral
\begin{equation}
t_{pd}\rightarrow t_{pd}-v_{pd}x^{2},\text{ \ \ where }v_{pd}>0,
\end{equation}
then to order $v_{pd}$, and defining $t=t_{pd}^{2}/\epsilon_{pd}$%
\begin{align}
\widetilde{H^{d}}  &  =\left(  \epsilon_{d}+2t\right)  \sum_{i,\sigma
}n_{i\sigma}-t\sum_{\left\langle i,j\right\rangle }X_{ij} \label{formulation2}%
\\
&  -\frac{2t_{pd}v_{pd}}{\epsilon_{pd}}\sum_{\left\langle i,j\right\rangle
,\sigma}\left(  n_{i\sigma}+n_{j\sigma}\right)  x_{ij}^{2}+\frac{2t_{pd}%
v_{pd}}{\epsilon_{pd}}\sum_{\left\langle i,j\right\rangle }X_{ij}x_{ij}%
^{2}.\nonumber
\end{align}

Restoring our original notation \cite{FBM1} $2t_{pd}v_{pd}/\epsilon
_{pd}=v/2\sqrt{nn_{s}}$, when the coupling $v$ is seen to be
\textbf{positive}
\begin{align}
\widetilde{H^{d}}  &  =\left(  \epsilon_{d}+2t\right)  \sum_{i,\sigma
}n_{i\sigma}-t\sum_{\left\langle i,j\right\rangle }X_{ij}\label{formulation2a}%
\\
&  -\frac{v}{2\sqrt{nn_{s}}}\sum_{\left\langle i,j\right\rangle ,\sigma
}\left(  n_{i\sigma}+n_{j\sigma}\right)  x_{ij}^{2}+\frac{v}{2\sqrt{nn_{s}}%
}\sum_{\left\langle i,j\right\rangle }X_{ij}x_{ij}^{2}.\nonumber
\end{align}

We retrieve our previous 1-band model (next-nearest and next-next-nearest
neighbor hoppings are dropped due to neglect of $t_{pp}$), but with an extra
term diagonal in $d$-space. As regards the vibrator, the effect of the new
term is to weaken the oxygen parabolic potential linearly with increasing
electron occupation of the band (or stiffen the vibrator with increasing hole
occupation). The number operator term is dominant over the hopping term (%
$<$%
X%
$>$
maximizes at $\simeq0.6$).

Let us now alternatively assume that the oxygen motion enters the $3$-band
hamiltonian through the interaction of the electrostatic potential with the
charge on the oxygen%
\begin{equation}
\epsilon_{pd}\rightarrow\epsilon_{pd}+v_{p}x^{2};
\end{equation}
where $v_{p}$ depends on a Madelung sum. In an ionic crystal it is arguable
that the sign of $\ v_{p}$ will be positive since the environment of a
negative ion typically consists of positive ions, so as the O-ion approaches
them the local oxide gap $\epsilon_{pd}$ becomes larger. However in a
perovskite structure the issue needs specific calculation.

Expanding to first order%
\begin{equation}
\frac{1}{\epsilon_{pd}+v_{p}x^{2}}=\frac{1}{\epsilon_{pd}}-\frac{v_{p}x^{2}%
}{\epsilon_{pd}^{2}}.
\end{equation}
Returning to Eq. (\ref{formulation1}), we insert the foregoing expansion into
the 2 terms to obtain
\begin{equation}
\Delta\widetilde{H^{d}}\rightarrow-\frac{tv_{p}}{\epsilon_{pd}}\sum
_{\left\langle i,j\right\rangle }\left(  n_{i\sigma}+n_{j\sigma}\right)
x_{ij}^{2}+\frac{tv_{p}}{\epsilon_{pd}}\sum_{\left\langle i,j\right\rangle
}X_{ij}x_{ij}^{2}. \label{formulation3}%
\end{equation}

The effect of the oscillator correction (\ref{formulation3}) from this
mechanism can be absorbed into (\ref{formulation2a}), giving the same final
result (\ref{formulation2a})\ but with
\begin{equation}
\frac{v}{2\sqrt{nn_{s}}}=\left(  2t_{pd}v_{pd}+tv_{p}\right)  /\epsilon_{pd}.
\end{equation}
The sign of $v$ will be positive if the $t_{pd}v_{pd}$ term in parenthesis is
dominant, or if $v_{p}$ is positive as argued above.

In this section we have formally derived the FBM coupling, showing the
approximations involved explicitly, and demonstrated the existence of a new
term in the coupling.

\section{FBM\ Hamiltonian}

The FBM\ Hamiltonian involves three pieces%
\begin{equation}
H=H^{v}+H^{e}+H^{ev}.
\end{equation}
In $H$ the Cu sites, which define the unit cell, are defined as 2D
integral-component vectors $\mathbf{i}=\left(  i_{x},i_{y}\right)  $ (lattice
constant is taken as unity). The two oxygens in each unit cell $\mathbf{i}$
are located at the sites $\mathbf{i+}\widehat{\mathbf{\alpha}}/2$, where
$\widehat{\mathbf{\alpha}}$ is a unit vector along the $x$- or $y$- axes,
hence $\widehat{\mathbf{\alpha}}$ defines whether the oxygen is in a Cu-O-Cu
bond oriented along the $x$- or $y$- direction.

In the vibrator piece $H^{v}$ the oxygen degree of freedom is an $n$-component
vector $\mathbf{x}_{\mathbf{i+}\widehat{\mathbf{\alpha}}/2}$, where $n=1$ if a
single mode is dominant (as assumed in the manuscript), $n=2$ if the two modes
transverse to the Cu-O-Cu bond are roughly equivalent, or in a case now
considered unlikely (as the along-bond mode is found to be weakly coupled)
$n=3$ if the two transverse modes and the along-bond mode can all be
considered equivalent. $H^{v}$ is given by
\begin{equation}
H^{v}=\sum_{\mathbf{i,\alpha=x}}^{\mathbf{y}}\left[  \frac{1}{2m}%
p_{\mathbf{i+}\widehat{\mathbf{\alpha}}/2}^{2}+\frac{\chi_{0}}{2}%
x_{\mathbf{i+}\widehat{\mathbf{\alpha}}/2}^{2}+\frac{w}{8n}\left(
x_{\mathbf{i+}\widehat{\mathbf{\alpha}}/2}^{2}\right)  ^{2}\right]  .
\label{vibrational}%
\end{equation}
In $H^{v}$ the scalar products $\mathbf{x}_{\mathbf{i+}\widehat{\mathbf{\alpha
}}/2}\cdot\mathbf{x}_{\mathbf{i+}\widehat{\mathbf{\alpha}}/2}$ are abbreviated
to $x_{\mathbf{i+}\widehat{\mathbf{\alpha}}/2}^{2}$, and a momentum
$\mathbf{p}_{\mathbf{i+}\widehat{\mathbf{\alpha}}/2}$\ conjugate to coordinate
$\mathbf{x}_{\mathbf{i+}\widehat{\mathbf{\alpha}}/2}$ is introduced, to define
the vibrator kinetic energy, with $m$ the oxygen mass ($M$ in the Ms.). The
"bare" bond force constant is $\chi_{0}$. The quartic term, with coefficient
$w$, is assumed in the degenerate case to be radially ($n=2$) or spherically
($n=3$) symmetric.

The electronic piece $H^{e}$ is
\begin{equation}
H^{e}=-\frac{1}{2}\sum_{\mathbf{i,j},\sigma}t\left(  \mathbf{i}-\mathbf{j}%
\right)  c_{\mathbf{i},\sigma}^{+}c_{\mathbf{j},\sigma}, \label{electronic}%
\end{equation}
where $c_{\mathbf{i},\sigma}^{+}$($c_{\mathbf{i},\sigma}$) denote respectively
the creation (destruction) operators for the $3d_{x^{2}-y^{2}}$ orbital (or,
more rigorously, the $d_{x^{2}-y^{2}}$-type Cu$3d$-O$2p$ antibonding Wannier
function) on lattice site $\mathbf{i}$ of spin $\sigma$. The strongest
interaction is the nearest neighbor hopping integral $t(\pm1,0)=t(0,\pm1)=t$,
($t$ is positive), followed by the next-nearest neighbor interaction
$t(\pm1,\pm1)=t^{\prime}$, ($t^{\prime}$ is negative) and then the 3rd-nearest
neighbor interaction $t(\pm2,0)=t(0,\pm2)=t^{\prime\prime}$ ($t^{\prime\prime
}$ is positive). The band eigenvalues $\epsilon_{\mathbf{k}}$ of
(\ref{electronic}) are
\begin{equation}
\epsilon_{\mathbf{k}}=-2t(\cos k_{x}+\cos k_{y})-4t^{\prime}\cos k_{x}\cos
k_{y}-2t^{\prime\prime}(\cos2k_{x}+\cos2k_{y}). \label{band_structure}%
\end{equation}
The model band structure has a minimum at $\Gamma$ ($\mathbf{k}=(0,0)$), a
maximum at $Z$ ($\mathbf{k}=(\pi,\pi)$), and saddle points (SP)\ at X
($\mathbf{k}=(0,\pi)$), and Y ($\mathbf{k}=(\pi,0)$). As a result of the
saddle points, located at $\epsilon_{\mathbf{SP}}=$ $4t^{\prime}%
-4t^{\prime\prime}$, the density of states (DOS) has a logarithmic peak (van
Hove singularity or vHs) at $\epsilon_{\mathbf{SP}}$ which is found from ARPES
and band structure calculations for near-optimally doped systems to lie close
to the Fermi level \cite{OKA2bs,ZX214} - the resulting high DOS at the Fermi
level strongly enhances the FBM coupling. The total band width is $8t$.

The electron-vibrator coupling piece is
\begin{align}
H^{ev}  &  =\frac{v}{2\sqrt{nn_{s}}}\sum_{\mathbf{i,\alpha=x}}^{\mathbf{y}%
}x_{\mathbf{i+}\widehat{\mathbf{\alpha}}/2}^{2}\left[  -\sum_{\sigma}\left(
n_{\mathbf{i,\sigma}}+n_{_{\mathbf{i+}\widehat{\mathbf{\alpha}}},\sigma
}\right)  +X_{\mathbf{i+}\widehat{\mathbf{\alpha}}/2}\right]  ;\text{
\ \ \ \ }\label{full_coupling}\\
X_{\mathbf{i+}\widehat{\mathbf{\alpha}}/2}  &  =\sum_{\sigma}\left(
c_{\mathbf{i},\sigma}^{+}c_{\mathbf{i+}\widehat{\mathbf{\alpha}},\sigma
}+c_{\mathbf{i+}\widehat{\mathbf{\alpha}},\sigma}^{+}c_{\mathbf{i},\sigma
}\right)  ,
\end{align}
where the bond order operator $X$ is associated with the oxygen site at the
bond center, and we have defined in the mixed degeneracy factor $\left(
nn_{s}\right)  ^{-1/2}$, where $n_{s}=2$ is the spin degeneracy, to make the
term of order $\sqrt{nn_{s}}$, motivated by a version of large-$N$ theory
jointly expanding in $1/n$ and $1/n_{s}$. In Ref. \cite{FBM1} only the
$X$-piece of (\ref{full_coupling}) was included.

The combination $-\sum_{\sigma}\left(  n_{\mathbf{i,\sigma}}+n_{_{\mathbf{i+}%
\widehat{\mathbf{\alpha}}},\sigma}\right)  +X_{\mathbf{i+}\widehat
{\mathbf{\alpha}}/2}$ can also be written in more compact form, defining the
antibonding orbital $\left\vert a,\mathbf{i+}\widehat{\mathbf{\alpha}%
}/2\right\rangle =\left(  \left\vert \mathbf{i}\right\rangle -\left\vert
\mathbf{i+}\widehat{\mathbf{\alpha}}\right\rangle \right)  /\sqrt{2}$, with
number operator $n_{\mathbf{i+}\widehat{\mathbf{\alpha}}/2}^{a}$ (summing over
spin). In the manuscript $n^{a}$ is simplified to $N$.%
\begin{equation}
-\sum_{\sigma}\left(  n_{\mathbf{i,\sigma}}+n_{\mathbf{i+\widehat
{\mathbf{\alpha}},\sigma}}\right)  +X_{\mathbf{i+}\widehat{\mathbf{\alpha}}%
/2}=-2n_{\mathbf{i+}\widehat{\mathbf{\alpha}}/2}^{a}. \label{interaction2}%
\end{equation}

The complete Hamiltonian $H=H^{v}+H^{e}+H^{ev}$ is then%
\begin{align}
H  &  =\sum_{\mathbf{i,\alpha=x}}^{\mathbf{y}}\left[  \frac{1}{2m}%
p_{\mathbf{i+}\widehat{\mathbf{\alpha}}/2}^{2}+\frac{\chi_{0}}{2}%
x_{\mathbf{i+}\widehat{\mathbf{\alpha}}/2}^{2}+\frac{w}{8n}\left(
x_{\mathbf{i+}\widehat{\mathbf{\alpha}}/2}^{2}\right)  ^{2}\right]  -\frac
{1}{2}\sum_{\mathbf{i,j},\sigma}t\left(  \mathbf{i}-\mathbf{j}\right)
c_{\mathbf{i},\sigma}^{+}c_{\mathbf{j},\sigma}\label{Hamiltonian}\\
&  +\frac{v}{\sqrt{nn_{s}}}\sum_{\mathbf{i,\alpha=x}}^{\mathbf{y}%
}x_{\mathbf{i+}\widehat{\mathbf{\alpha}}/2}^{2}n_{\mathbf{i+}\widehat
{\mathbf{\alpha}}/2}^{a}.\nonumber
\end{align}
Note that in Eq.(\ref{Hamiltonian}) $K=v^{2}/w$ defines a coupling energy.

\section{Determination of Coupling $v$}

Calculation of the oxygen PE surface as a function of doping is not an ideal
approach to calculationg the FBM coupling constant for two reasons. The PE
surface calculations such as illustrated in Fig. 2 in the Ms. are relaxed
surfaces, i.e. when certain oxygen coordinates are held fixed all other atoms
are allowed to find their minumum energy. This differs from the "vertical"
couplings entering the FBM Hamiltonian, where atoms other than planar oxygen
are assumed fixed in place. Secondly the coupling in the FBM Hamiltonian is to
the number of electrons $n^{a}$ in the antibonding orbital, which mainly
involves states at the top of the $d$-band and will be filled mainly by adding
electrons rather than holes as was done (for reasons of computational
stability) in Fig. 2 of the Ms..

The method adopted to calculate the coupling strength $v$ is based on
comparing the shift in band structure energies when the oxygen location is
perturbed with the same shift deduced from the FBM Hamiltonian. The
FBM\ coupling (third term in Eq.(\ref{Hamiltonian})) leads to splittings in
the tight-binding band structure. If all oxygens in the $x$-oriented bonds are
globally shifted by $u_{x}$, and all oxygens in the $y$-oriented bonds by
$u_{y}$, there is a splitting of the band energy between the band energy
$\epsilon_{X}$ at the saddle point (SP) X$=\left(  \pi,0\right)  $, and
$\epsilon_{Y}$ at Y$=\left(  0,\pi\right)  $, given by $\epsilon_{X}%
-\epsilon_{Y}=\sqrt{2/n}v\left(  u_{x}^{2}-u_{y}^{2}\right)  .$ By numerically
calculating the band structure with first the $x$-oxygens displaced, and then
the $y$-oxygens, and subtracting the corresponding band structure energies
energies at, say, the SP X, any isotropic shift resulting from displacing a
single oxygen can be cancelled out and the coupling $v$ determined. The
results are shown in Table I.

\section{Mean Field Approximation}

Mean field theory is a useful step in investigating the properties of many
models. In the FBM, the mean field approximation decouples the electronic and
vibrational parts of the Hamiltonian. In the vibrational part, an expectation
value of the electronic terms shifts the oscillator harmonic frequency, the
expectation value being assumed spatially uniform, but it can be different in
the $x$- and $y$- bonds (in this section we return to the notation in the
Ms.):%
\begin{align}
H^{vib}  &  =\sum_{\left\langle i,j\right\rangle }\frac{p_{ij}^{2}}{2M}%
+\frac{1}{2}\sum_{\left\langle i,j\right\rangle }\chi_{0}u_{ij}^{2}+\frac
{w}{8}\sum_{\left\langle i,j\right\rangle }u_{ij}^{4}
\label{effective_quartic}\\
&  +\frac{v}{2\sqrt{2}}\sum_{\left\langle i,j\right\rangle ,\sigma}\left(
2-2p+\left\langle c_{i,\sigma}^{+}c_{j,\sigma}+c_{j,\sigma}^{+}c_{i,\sigma
}\right\rangle \right)  u_{ij}^{2}.\nonumber
\end{align}
$H^{vib}$ can easily be diagonalized in a harmonic oscillator basis. In the
electronic part, the expectation value of the square of the oscillator
amplitude has been taken,%
\begin{equation}
H^{el}=\sum_{\mathbf{k},\sigma}\epsilon_{\mathbf{k}}n_{\mathbf{k},\sigma
}+\frac{v}{2\sqrt{2}}\sum_{\left\langle i,j\right\rangle ,\sigma}\left[
c_{i,\sigma}^{+}c_{j,\sigma}+c_{j,\sigma}^{+}c_{i,\sigma}\right]  \left\langle
u_{ij}^{2}\right\rangle , \label{effective_el}%
\end{equation}
giving a band structure problem in which there are new nearest-neighbor
hopping terms $\left(  v/2\sqrt{2}\right)  \left[  c_{i,\sigma}^{+}%
c_{j,\sigma}+c_{j,\sigma}^{+}c_{i,\sigma}\right]  \left\langle u_{ij}%
^{2}\right\rangle $ (the uniform shift represented by the number operator
terms does not change the band structure and is omitted) with the effect of
reducing the nearest-neighbor hopping integral. Allowing the oscillator
amplitude squared for the $x$-directed $\left\langle u_{ij}^{2}\right\rangle
_{x}$ and $y$-directed $\left\langle u_{ij}^{2}\right\rangle _{y}$\ bonds to
be unequal (the C4 symmetry-split case), the band structure is changed to
\begin{equation}
\widetilde{\epsilon}_{\mathbf{k}}=\epsilon_{\mathbf{k}}+\frac{v}{\sqrt{2}%
}\left\langle u_{ij}^{2}\right\rangle _{x}\cos k_{x}+\frac{v}{\sqrt{2}%
}\left\langle u_{ij}^{2}\right\rangle _{y}\cos k_{y}. \label{eff_bs}%
\end{equation}

Using the band structure $\widetilde{\epsilon}_{\mathbf{k}}$ (\ref{eff_bs})
the expectation values $\left\langle c_{i,\sigma}^{+}c_{j,\sigma}+c_{j,\sigma
}^{+}c_{i,\sigma}\right\rangle $ for $x$-oriented and $y$-oriented bonds are
calculated, hence defining two quartic Hamiltonians (\ref{effective_quartic})
whose exact solution yields the squared vibrator amplitudes $\left\langle
u_{ij}^{2}\right\rangle _{x}$ and $\left\langle u_{ij}^{2}\right\rangle _{y}$.
These interconnected electronic and quartic problems are then solved
self-consistently as regards the expectation values. The parameters used were
similar to Table I, $v=0.0198$ au, $w=0.085$ au, the oscillator bare force
constant was $\chi_{0}=-0.0225$ au. The band structure is parametrized by the
(negative of the) hopping matrix elements, the nearest-neighbor hopping matrix
element $t=0.25$ eV, next-nearest-neighbor hopping m.e. $t^{\prime}=-0.05$ eV,
and third next-nearest-neighbor hopping m.e. $t^{\prime\prime}=27.2$ meV.

We can rewrite the effective band structure as
\begin{equation}
\widetilde{\epsilon}_{\mathbf{k}}=\overline{\epsilon_{\mathbf{k}}}+\frac{1}%
{2}\Delta_{ps}\left(  \cos k_{x}-\cos k_{y}\right)  ,
\end{equation}
where $\Delta_{ps}=\left(  v/\sqrt{2}\right)  \left(  \left\langle u_{ij}%
^{2}\right\rangle _{x}-\left\langle u_{ij}^{2}\right\rangle _{y}\right)  $ is
the pseudogap, and the renormalized nearest-neighbor hopping $\left(
v/2\sqrt{2}\right)  \left(  \left\langle u_{ij}^{2}\right\rangle
_{x}+\left\langle u_{ij}^{2}\right\rangle _{y}\right)  $ is absorbed into
$\overline{\epsilon_{\mathbf{k}}}$. $\ $The experimental data \cite{Kohsaka1}
show that the pseudogap is not uniform over the sample as we have, for
simplicity, assumed, but the coherence length over which the sign of
$\Delta_{ps}^{0}$ varies is quite short, only a few lattice spacings. Probably
as a result of this nanoscopic domain structure, the phase boundary of the
pseudogap region is not typically found experimentally to constitute a true,
sharp, phase boundary \cite{PGReview}.

The variation of pseudogap with doping at low temperature seen in the contour
plot (Manuscript Fig. 3) is similar to thet seen in experimental data
\cite{JinhoLee} (see Fig. 2)%
\begin{figure}
[ptb]
\begin{center}
\includegraphics[
trim=3.051982in 1.740398in 0.657645in 0.000000in,
height=2.9698in,
width=3.2188in
]%
{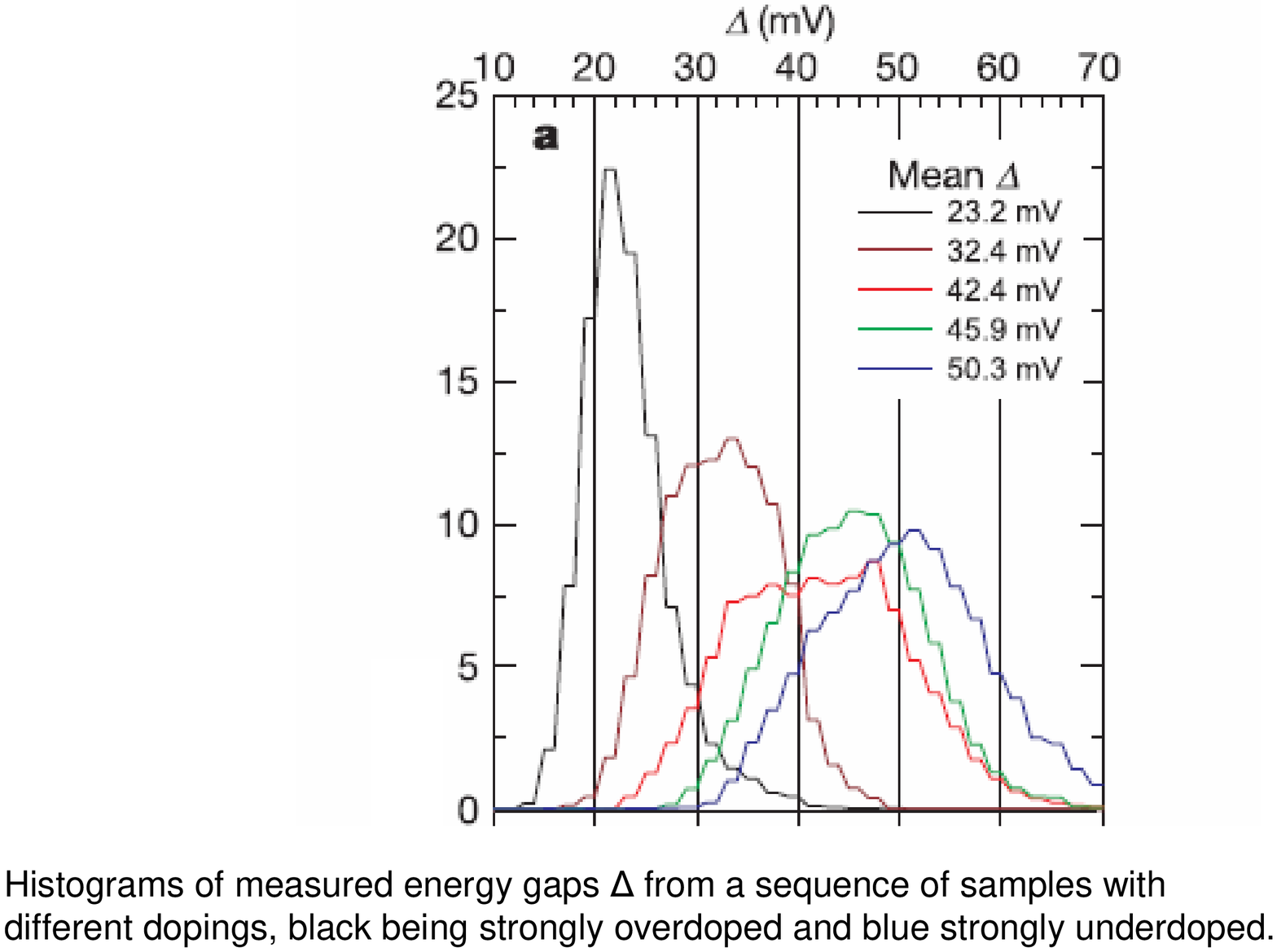}%
\caption{Histograms of measured energy gaps $\Delta$ from a sequence of
samples with different dopings, black being strongly overdoped and blue
strongly underdoped \cite{JinhoLee}.}%
\end{center}
\end{figure}

\section{Intensity Variation in Experimental $R$-plots}

In order to model the experimental behavior in the STM experiments
\cite{Kohsaka1} on C4 symmetry-split systems, we calculated the projected DOS
for a 3-band model with the basis of oxygen $2p_{x}$ and $2p_{y}$ orbitals and
Cu $3d_{x^{2}-y^{2}}$ orbitals shown in Fig. 1. The $pd\sigma$ hopping matrix
element is $t_{pd}=1.12$ eV. \ There are $pp$ hopping matrix elements between
nearest-neighbor $2p_{x}$ and $2p_{y}$ orbitals given by $t_{pp}=-0.528$ eV,
and an oxide gap $\epsilon_{d}-\epsilon_{p}=6$ eV. A spatially-uniform
pseudogap is introduced by modifying the $t_{pd}$ matrix elements to
$t_{p_{x}d}=t_{pd}+\Delta t$ (i.e. for the lower vibrational amplitude oxygen)
and $t_{p_{y}d}=t_{pd}-\Delta t$ (i.e. for the higher vibrational amplitude
oxygen), where $\Delta t=0.0375$ eV (the argument below only depends on these
being semiquantitatively correct). The results for the DOS projected into the
oxygen $2p_{x}$ orbitals (lying in $x$-oriented Cu-O-Cu bonds - see Fig. 1)
and oxygen $2p_{y}$ orbitals are different, as seen in Fig. 3. The DOS peak
associated with the van Hove singularity is seen in Fig. 3 to be split, the
peak above the Fermi level being localized only on the lower vibrational
amplitude oxygen, and the peak below the Fermi level being being localized
only the higher vibrational amplitude oxygen. The STM $R$-map technique
\cite{Kohsaka1} for detecting the C4 splitting experimentally involves the
ratio $R$ of the tunneling current into the empty DOS to the hole current into
the filled DOS. Evidently from Fig3 , $R$ is predicted to be large on the low
amplitude oxygens and small on the high amplitude oxygens, in agreement with
the observation \cite{Kohsaka1}, in which the high amplitude oxygens are
associated with dark streaks in the $R$-map, while the low amplitude oxygens
are associated with bright spots. Note that the C4 splitting is characterized
by nanoscale domains \cite{Kohsaka1}.%
\begin{figure}
[ptb]
\begin{center}
\includegraphics[
trim=1.736693in 1.083276in 2.837024in 1.097101in,
height=2.7691in,
width=2.8193in
]%
{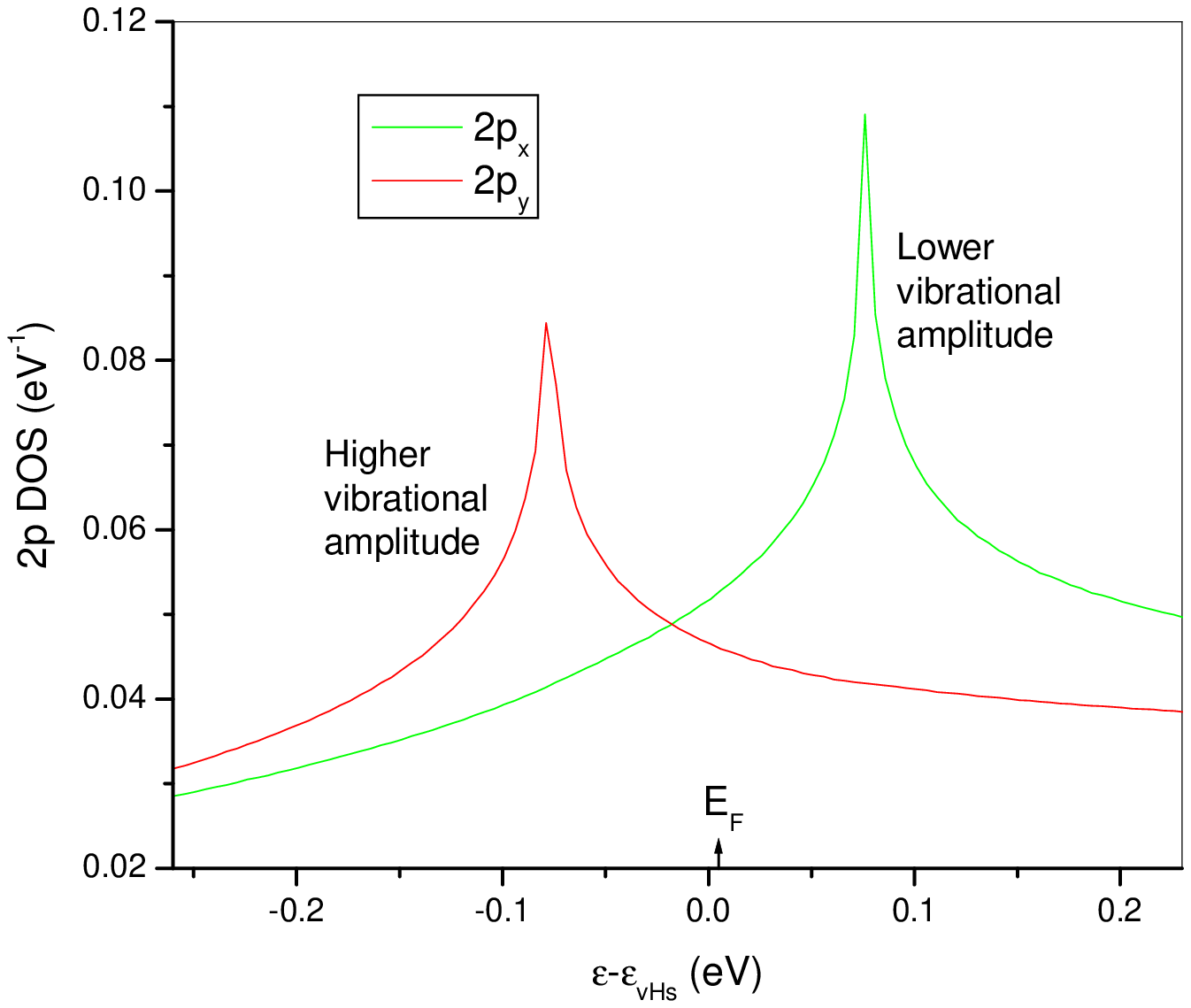}%
\caption{Oxygen projected $2p$-DOS for oxygens in $x$-oriented and
$y$-oriented bonds.The peak above the Fermi level is for the lower vibrational
amplitude oxygen, and the peak below the Fermi level is for the higher
vibrational amplitude oxygen.}%
\end{center}
\end{figure}

\section{Lagrangian Formalism}

Here we wish to derive the pairing interaction in the FBM. For this purpose a
simple approach, akin to the familiar RPA, is to utilise the $1/N$ expansion
technique, where $N$ is degeneracy, e.g. vibrational degeneracy $n$ or spin
degeneracy $n_{s}$. In this approach the exact solution to the $n=1$ quartic
vibrator used in the main Ms. will be replaced by the (very similar) mean
field solution to the $n>1$ quartic vibrator \cite{FBM1}. The $1/N$ expansion
works well e.g. for the Kondo problem \cite{Read-Newns}, the results remaining
physical down to spin degeneracy $n_{s}=2$. Here we symetrically co-expand in
the inverse of the mode degeneracy $n$ and the spin degeneracy $n_{s}$
\cite{Coleman}, meaning by expressions such as "$1/N$" the joint orders
\thinspace$1/n$ and $1/n_{s}$. The $1/N$ expansion technique is usually
implemented within a Lagrangian/path integral formulation \cite{Coleman}, the
approach we shall adopt here. We take $\hbar=1$, and omit the sum-normalizing
factors $1/N_{x}^{2}$, which can always be restored by inspection.

Within the usual imaginary time (Euclidean) Lagrangian formulation at finite
temperature the partition function can be written
\begin{equation}
Z=\int\mathcal{D}x\mathcal{D}ce^{-\int_{0}^{\beta}\mathcal{L}\left(
\tau\right)  d\tau},
\end{equation}
where $\mathcal{L}\left(  \tau\right)  $ is the Lagrangian as a function of
imaginary time $\tau$, $\beta$ is $1/T$ ($T$ is temperature, taking
Boltzmann's constant $k_{B}=1$), $\int\mathcal{D}x\mathcal{D}c$ implies a path
integral over the vibrator and fermionic (Grassman) variables.

The Lagrangian, readily derived from the Hamiltonian (\ref{Hamiltonian}),
\begin{equation}
\mathcal{L}=\mathcal{L}_{e}^{(1)}+\mathcal{L}_{v}^{(1)}+\mathcal{L}_{v}%
^{(2)}+\mathcal{L}_{ev}^{(2)}, \label{Lagrangian}%
\end{equation}
comprises the previously described terms (\ref{electronic}),
(\ref{vibrational}), and (\ref{full_coupling})
\begin{align}
\mathcal{L}_{e}^{(1)}  &  =\sum_{\mathbf{i,}\sigma}c_{\mathbf{i},\sigma}%
^{+}\frac{\partial}{\partial\tau}c_{\mathbf{i,\sigma}}-\frac{1}{2}%
\sum_{\mathbf{i,j},\sigma}t\left(  \mathbf{i}-\mathbf{j}\right)
c_{\mathbf{i},\sigma}^{+}c_{\mathbf{j},\sigma},.\\
\mathcal{L}_{v}^{(1)}  &  =\frac{1}{2}\sum_{\mathbf{i,\alpha}}\left[  m\dot
{x}_{\mathbf{i+}\widehat{\mathbf{\alpha}}/2}^{2}+\chi_{0}x_{\mathbf{i+}%
\widehat{\mathbf{\alpha}}/2}^{2}\right]  ,\nonumber\\
\mathcal{L}_{v}^{(2)}  &  =\frac{w}{8n}\sum_{\mathbf{i,\alpha}}\left(
x_{\mathbf{i+}\widehat{\mathbf{\alpha}}/2}^{2}\right)  ^{2},\nonumber\\
\mathcal{L}_{ev}^{(2)}  &  =\frac{v}{\sqrt{nn_{s}}}\sum_{\mathbf{i,\alpha=x}%
}^{\mathbf{y}}x_{\mathbf{i+}\widehat{\mathbf{\alpha}}/2}^{2}n_{\mathbf{i+}%
\widehat{\mathbf{\alpha}}/2}^{a}\nonumber
\end{align}
A kinetic energy for the fermion variables is introduced on line 1, otherwise
the terms are the same as in Eq's (\ref{electronic}), (\ref{vibrational}), and
(\ref{full_coupling}). In the following we extend the earlier FBM results
\cite{FBM1} to include the more complete form of coupling derived in Sec. II.

\section{Results}

We implement a Stratonovic decoupling on the bilinear terms $\mathcal{L}%
_{v}^{(2)}$ and $\mathcal{L}_{ev}^{(2)}$ in the Lagrangian, generating path
integrals over new fields $z_{v,\mathbf{\mathbf{i+}\widehat{\mathbf{\alpha}%
}/2}}$ and $z_{e,\mathbf{\mathbf{i+}\widehat{\mathbf{\alpha}}/2}}$ defined on
each Cu-O-Cu bond. Later this will make it possible to implement the path
integrals over the Boson and Fermion fields under the $z$-path integral.

Now the partition function is written in terms of the action $S$%
\begin{equation}
Z=\int\mathcal{D}z\mathcal{D}x\mathcal{D}ce^{-\beta\mathcal{S}},
\end{equation}
which is again decomposed into the terms
\begin{equation}
S=S_{v}^{(1)}+S_{e}^{(1)}+S_{z}^{(2)}+S_{zv}+S_{ze}, \label{Action_full}%
\end{equation}
defined by
\begin{align}
S_{v}^{(1)}  &  =\frac{m}{2}\sum_{\mathbf{k},m,\alpha}\left(  \omega_{m}%
^{2}+\overline{\omega}_{\alpha}^{2}\right)  x_{-\mathbf{k},-m,\alpha
}x_{\mathbf{k},m,\alpha},\\
S_{e}^{(1)}  &  =\sum_{\mathbf{k},m,\sigma}\left(  \overline{\epsilon
}_{\mathbf{k}}-i\nu_{m}\right)  c_{\mathbf{k},m,\sigma}^{+}c_{\mathbf{k}%
,m,\sigma},\nonumber\\
S_{z}^{(2)}  &  =\frac{n_{s}}{2K}\sum_{\mathbf{q},n,\alpha}z_{-\mathbf{q}%
,-n,\alpha}^{e}z_{\mathbf{q},n,\alpha}^{e}-\frac{2\sqrt{nn_{s}}}{v}%
\sum_{\mathbf{q},\alpha,n}z_{-\mathbf{q},-n,\alpha}^{e}z_{\mathbf{q},n,\alpha
}^{v},\nonumber\\
S_{zv}  &  =-\sum_{\mathbf{q,k},n,m,\alpha}z_{-\mathbf{q},-n,\alpha}%
^{v}x_{-\mathbf{k},-m,\alpha}x_{\mathbf{k+q},m+n,\alpha}+\sum_{\alpha
}z_{\mathbf{0},0,\alpha}^{v}\left\langle x^{2}\right\rangle ,\nonumber\\
S_{ze}  &  =\sum_{\mathbf{q},n,\sigma,\alpha}\psi_{\mathbf{k,k}+\mathbf{q}%
}^{\alpha}z_{-\mathbf{q},-n,\alpha}^{e}c_{\mathbf{k},m,\sigma}^{+}%
c_{\mathbf{k+q},m+n,\sigma}-2\sum_{\alpha}z_{\mathbf{0},0,\alpha}%
^{e}\left\langle n^{a}\right\rangle .\nonumber
\end{align}
Here%
\begin{align}
x(t)  &  =\sum_{n}e^{-i\omega_{n}t}x_{n},\text{ \ }\omega_{n}=2n\pi T;\text{
\ \ \ \ \ \ \ \ \ \ Bose and z fields,}\\
c(t)  &  =\sum_{n}e^{-i\nu_{n}t}c_{n},\text{ \ }\nu_{n}=\left(  2n+1\right)
\pi T;\text{ \ \ Fermion fields,}\nonumber
\end{align}
define the Fourier series converting to Bosonic and Fermionic Matsubara
frequencies $\omega_{n}$ and $\nu_{n}$, and a coupling amplitude
$\psi_{\mathbf{k,k}^{\prime}}^{\alpha}$ different from that in \cite{FBM1}
appears
\begin{equation}
\psi_{\mathbf{k,k}^{\prime}}^{\alpha}=4\sin\left[  k_{\alpha}/2\right]
\sin\left[  k_{\alpha}^{\prime}/2\right]  . \label{chi}%
\end{equation}
The band structure $\overline{\epsilon}_{\mathbf{k}}$ and vibrator frequencies
$\overline{\omega}_{\alpha}$ now involve mean field quantities, an effective
nearest-neighbor hopping integral $t_{\alpha}$ (\ref{t_alpha}) defined by%
\begin{equation}
t_{\alpha}=t+z_{e}^{\alpha}=t-\frac{v}{2\sqrt{nn_{s}}}\left\langle
x^{2}\right\rangle _{\alpha}, \label{t_alpha}%
\end{equation}
physically meaning that the effective hopping is weakened by the vibrational
displacement of the oxygen, and vibration frequencies $\overline{\omega
}_{\alpha}$ given by (\ref{freq2})%
\begin{align}
m\overline{\omega}_{\alpha}^{2}  &  =\chi_{0}-2z_{v}^{\alpha}\label{freq2}\\
&  =\chi_{0}-\frac{2v}{\sqrt{nn_{s}}}\left\langle n^{a}\right\rangle _{\alpha
}+\frac{w}{2n}\left\langle x^{2}\right\rangle _{\alpha}.\nonumber
\end{align}
physically meaning that the quasiharmonic vibrator is softened by the number
of antibonding electrons $\left\langle n^{a}\right\rangle _{\alpha}$, and
stiffened by the effect of the quartic potential.

On implementing the path integrals over the Boson and Fermion fields under the
$z$-path integral (here we drop the $\alpha$-dependence of the mean field
quantities for simplicity), we obtain to order $(1/N)$ the quasi-harmonic
action
\begin{equation}
S=\frac{1}{2}\sum_{\mathbf{q},n}\left[  z_{-\mathbf{q},-n,x}^{v}%
,z_{-\mathbf{q},-n,y}^{v}z_{-\mathbf{q},-n,x}^{e},z_{-\mathbf{q},-n,y}%
^{e}\right]  \mathbf{A}(\mathbf{q},n)\left[
\begin{array}
[c]{c}%
z_{\mathbf{q},n,x}^{v}\\
z_{\mathbf{q},n,y}^{v}\\
z_{\mathbf{q},n,x}^{e}\\
z_{\mathbf{q},n,y}^{e}%
\end{array}
\right]  ,
\end{equation}
where the matrix $\mathbf{A}$ is given by%
\begin{equation}
\mathbf{A}(\mathbf{q},n)=\left[
\begin{array}
[c]{cccc}%
-nD_{2}(n) & 0 & -\frac{2\sqrt{nn_{s}}}{v} & 0\\
0 & -nD_{2}(n) & 0 & -\frac{2\sqrt{nn_{s}}}{v}\\
-\frac{2\sqrt{nn_{s}}}{v} & 0 & \frac{n_{s}}{K}\left(  1-KR_{xx}%
(\mathbf{q},n)\right)  & -n_{s}R_{xy}(\mathbf{q},n)\\
0 & -\frac{2\sqrt{nn_{s}}}{v} & -n_{s}R_{yx}(\mathbf{q},n) & \frac{n_{s}}%
{K}\left(  1-KR_{yy}(\mathbf{q},n)\right)
\end{array}
\right]  ,
\end{equation}
and the Response functions (RF) are defined by
\begin{equation}
D_{2}\left(  \mathbf{q},n\right)  =\frac{2}{m^{2}\overline{\omega}}\left[
\frac{1}{\left(  \omega_{n}^{2}+4\overline{\omega}^{2}\right)  }\coth\left(
\frac{\overline{\omega}}{2T}\right)  +\frac{\delta_{n,0}}{8T\overline{\omega
}\sinh^{2}\left(  \frac{\overline{\omega}}{2T}\right)  }\right]  ,
\label{2phonon}%
\end{equation}
and ($f(\epsilon)$ is the Fermi function)%
\begin{equation}
R_{\alpha\beta}(\mathbf{q},n)=-\sum_{\mathbf{k}}\frac{f(\epsilon_{\mathbf{k}%
})-f(\epsilon_{\mathbf{k+q}})}{\epsilon_{\mathbf{k}}-\epsilon_{\mathbf{k+q}%
}+i\omega_{n}}\psi_{\mathbf{k,k+q}}^{\alpha}\psi_{\mathbf{k+q,k}}^{\beta}.
\end{equation}

\section{Discussion}

The important regions of $k$-space in the Brillouin Zone are the high-density
of states regions around the saddle points (SP)\ at $\mathbf{k}=(0,\pi)$
(termed Y), and $(\pi,0)$ (termed X). The key issue is to understand how the
FBM behaves at the SP. Because of the $\psi$-factors, the RF are close to
being diagonal if we assume only regions around the SP are important in the
$k$-sum. Therefore the $A$-matrix separates into a component having a pole in
the $x$-channel, and a component having a pole in the $y$-channel. At zero
frequency and wavevector, these poles correspond to the onset of the phase
boundary for the PG at $T^{\ast}$. As regards pairing, we consider the
interaction for a pair to scatter from $(\mathbf{k,-k})$ to $(\mathbf{k}%
^{\prime}\mathbf{,-k}^{\prime})$, which involves the factor $\left(
\psi_{\mathbf{k,k}^{\prime}}^{\alpha}\right)  ^{2}$. This will either scatter
where $\mathbf{k}$ and $\mathbf{k}^{\prime}$ lie in the neighborhood of the
point X which involves the factor $\left(  \psi_{\mathbf{k,k}^{\prime}}%
^{x}\right)  ^{2}\sim16$, in which case the inverse of the $A$-matrix in the
$x$-channel is involved, or in the neighborhood of the point Y, which involves
the factor $\left(  \psi_{\mathbf{k,k}^{\prime}}^{y}\right)  ^{2}\sim16$, in
which case the inverse of the $A$-matrix in the $y$-channel is involved. The
interaction has the classic form required for $d$-wave pairing, in that the
interaction is attractive (in fact close to singular) for $(\mathbf{k-k}%
^{\prime})$ small, and at large $(\mathbf{k-k}^{\prime})$ the attraction dies
off (in fact, because of the short-range repulsive interaction $U$, which is
flat in $k$-space, the interaction actually becomes repulsive at large
$(\mathbf{k-k}^{\prime})$).

The situation for the full FBM Hamiltonian (\ref{Hamiltonian}) is similar but
different in detail for the interaction which considers only the $X$-term in
the electronic coupling (\ref{interaction2}) \cite{FBM1}, when instead of the
coupling amplitude $\psi_{\mathbf{k,k}^{\prime}}^{\alpha}$ we have the
amplitude%
\begin{equation}
\chi_{\mathbf{k,k}^{\prime}}^{\alpha}=-2\cos\left(  \frac{k_{\alpha}%
+k_{\alpha}^{\prime}}{2}\right)  .
\end{equation}
This function leads to a different $A$-matrix \cite{FBM1}, in which only a
component with $d$-symmetry (the $z$'s in the $x$- and $y$- channels having
opposite sign) is singular, the other, $s$-channel, being nonsingular. Now
scattering around either SP X or SP Y is possible with the the factor $\left(
\chi_{\mathbf{k,k}^{\prime}}^{\alpha}\right)  ^{2}\sim4$ for either $\alpha=x$
or $\alpha=y$. However only one combination of channels is singular, so we
assume that this can be considered as if there is only one effective channel
for scattering around each SP, just as in the case where both terms in the
electronic coupling (\ref{interaction2}) are present. Hence we conclude on the
basis of dominance by the SP, that the pairing interaction in the case of the
full electronic coupling (\ref{interaction2}) is $4\times$ as large as that in
the reduced interaction considered earlier. This is the result cited in the
manuscript, which however needs verification by numerical solution of the gap equation.